\documentclass{article}

\begin{document}



\title{INTEGRABLE COUPLING IN A MODEL FOR JOSEPHSON TUNNELING 
BETWEEN NON-IDENTICAL BCS SYSTEMS
}

\author{Jon Links\footnote{
email: jrl@maths.uq.edu.au.} \\ 
Department of Mathematics, The University of Queensland, \\
Brisbane, 4072, Australia, \\
~~\\
Katrina E. Hibberd\footnote{email:
keh@dftuz.unizar.es} \\   
Departamento de F\'{\i}sica Te\'orica, Universidad de Zaragoza,\\
Zaragoza, Spain. 
}

\maketitle


\begin{abstract}
We extend a recent construction for an integrable model describing 
Josephson tunneling between identical BCS systems to the case where 
the BCS systems have different single particle energy levels.  
The exact solution of this generalized  model is obtained through the 
Bethe ansatz.  
\end{abstract}


\def\a{\alpha}
\def\b{\beta}
\def\d{\dagger}
\def\e{\epsilon}
\def\g{\gamma}
\def\k{\kappa}
\def\l{\lambda}
\def\o{\omega}
\def\t{\tilde{\tau}}
\def\s{S}
\def\D{\Delta}
\def\T{{\cal T}}
\def\TT{{\tilde{\cal T}}}

\def\beq{\begin{equation}}
\def\eeq{\end{equation}}
\def\bea{\begin{eqnarray}}
\def\eea{\end{eqnarray}}
\def\ba{\begin{array}}
\def\ea{\end{array}}
\def\no{\nonumber}
\def\le{\langle}
\def\re{\rangle}
\def\lt{\left}
\def\rt{\right}
\def\oR{R^*} \def\upa{\uparrow}
\def\R{\overline{R}} \def\doa{\downarrow}
\def\oL{\overline{\Lambda}}
\def\nn{\nonumber} \def\dag{\dagger}
\def\e{\epsilon}
\def\si{\sigma}
\def\th{\theta}
\def\de{\delta}
\def\ga{\gamma}
\def\l{\left}
\def\r{\right}
\def\a{\alpha}
\def\b{\beta}
\def\g{\gamma}
\def\La{\Lambda}
\def\w{\overline{w}}
\def\u{\overline{u}}
\def\o{\overline}
\def\rr{\mathcal{R}}
\def\T{\mathcal{T}}
\def\N{\overline{N}}
\def\Q{\overline{Q}}
\def\L{\mathcal{L}}
\def\k{\overline{k}}
\def\l{\overline{l}}
\def\d{\dagger}

\section{Introduction.}

The experimental work of Ralph, Black and Tinkham\cite{brt,rbt} 
on the discrete 
energy spectrum in small metallic aluminium grains has generated substantial
interest in understanding the nature of superconducting correlations at the 
nano-scale level. Their results indicate significant parity effects due to the 
number of electrons in the system. For grains with an odd number of electrons,
the gap in the energy spectrum reduces with the size of the system, in contrast 
to the case of a grain with an even number of electrons, where a gap 
larger than the single electron energy levels persists. In the latter case 
the gap can be closed by a strong applied magnetic field. 
The conclusion drawn from these results is that pairing interactions
are prominent in these nano-scale systems.
For a grain with an odd number of electrons there will always 
be at least one unpaired electron, so it is not necessary to break a Cooper
pair in order to create an excited state. For a grain with an even number 
of electrons, all excited states have a least one broken Cooper pair,
resulting in a gap in the spectrum.
In the prescence of a 
strongly applied magnetic field, it is energetically more favourable 
for a grain with an even number of electrons 
to have broken pairs, and hence in this case 
there are excitations which show no gap in the 
spectrum. 
    
The physical properties of a small metallic grain are described by the
reduced BCS Hamiltonian\cite{dr01,vd}   
\beq H_{BCS}=\sum_{j=1}^{\L}\e_jn_j
-g\sum_{j,k}^{\L}c_{k+}^{\d}c_{k-}^{\d}c_{j-}c_{j+}. \label{bcs}
\eeq 
Above, $j=1,...,{\L}$ labels a shell of doubly degenerate single particle
energy levels with energies $\e_j$ and $n_j$ is the 
fermion number operator for
level $j$. The operators $c_{j\pm},\,c^{\d}_{j\pm}$ are the annihilation
and creation operators for the fermions at level $j$. The labels $\pm$ refer 
to time reversed states.  

One of the prominent features of the Hamiltonian (\ref{bcs}) 
is the {\it blocking
effect}. For any unpaired electron at level $j$ the action of 
the pairing interaction is zero since only paired electrons are
scattered. This means that the Hilbert space can be decoupled into 
a product of paired and unpaired electron states in which the
action of the Hamiltonian on the subspace for the unpaired electrons is 
automatically diagonal in the natural basis. 
In view of the blocking effect, it is convenient to introduce 
hard-core boson operators 
$b_j=c_{j-}c_{j+},\, b^{\d}_{j}=c^{\d}_{j+}c^{\d}_{j-}$ which satisfy
the relations 
$$(b^{\d}_j)^2=0, ~~[b_j,\,b_k^{\d}]=\delta_{jk}(1-2b^{\d}_jb_j) $$
$$[b_j,\,b_k]=[b^{\d}_j,\,b^{\d}_k]=0 $$
on the subspace excluding single particle states. 
In this setting the hard-core boson operators realise the
$su(2)$ algebra in the pseudo-spin reprepresentation, 
which will be utilized below.

The original approach of Bardeen, Cooper and Schrieffer\cite{bcs} 
to describe the 
phenomenon of superconductivity was to employ
a mean field theory using a variational wavefunction for the ground state 
which has an undetermined number of electrons. The expectation value for the 
number operator is then fixed by means of a chemical potential term $\mu$. 
One of the 
predictions of the BCS theory is that the number of Cooper pairs in the ground 
state of the system is given by the ratio $\Delta/d$ where $\Delta$ is the BCS 
``bulk gap'' and $d$ is the mean level spacing for the single electron 
eigenstates. For nano-scale systems, this ratio is of the order of unity, in 
seeming contradiction with the experimental results discussed above. 
The explanation for this is that the mean-field approach is inappropriate
for nano-scale systems due to large superconducting fluctuations.

As an alternative to the BCS mean field approach, one can appeal to the 
exact solution of the Hamiltonian (\ref{bcs}) derived by Richardson 
and Sherman\cite{r63}.
It has also been shown by Cambiaggio, Rivas and Saraceno\cite{crs} that 
(\ref{bcs})
is integrable in the sense that there exists a set of mutually commutative 
operators which commute with the Hamiltonian. These features have recently been 
shown to be a consequence of the fact that the model can be derived in the 
context of the Quantum Inverse Scattering Method (QISM) using a solution 
of the Yang-Baxter equation associated with the Lie algebra 
$su(2)$\cite{zlmg,vp}.  
One of the aims of the present work is to extend this approach for application
to generalised models. As a specific example, 
we will show that a model for strong 
Josephson coupling between two BCS systems falls into this class. 

Recall first that electron pairing interactions manifest themselves in 
macroscopic systems via three well known phenomena: 
\begin{itemize} 
\item supercurrents 
\item Meissner effect 
\item Josephson effect 
\end{itemize} 
As noted by von Delft\cite{vd}, the notion of a supercurrent in a 
nano-scale system is inapplicable because the mean fee path of 
an electron is comparable to the system size. Likewise, the penetration 
depth of 
an applied magnetic field is comparable to the system size, which prohibits
any Meissner effect. 

Josephson\cite{j} put forth a proposal for the tunneling of electron pairs 
between superconductors separated by an insulating barrier. 
A theory was derived to describe {\it weak} coupling 
between two superconductors treated at the mean field level in the 
grand-canonical ensemble. A remarkable prediction of the theory 
was that it is possible for a direct current to flow across the 
insulator for the case of zero applied voltage, whereas a constant voltage 
across the insulator produces an alternating current. The essential 
features of the theory stem from the phase difference between the 
superconductors, which is well defined since the variational wavefunctions 
for the superconductors have undetermined particle numbers.     

For the case of nano-scale systems, the above 
predictions are  again invalid due to the finite particle numbers for each 
system, giving rise to phase uncertainty. 
However, if we are to consider {\it strong} coupling where individual particle 
numbers are not conserved, only total particle number, it is appropriate to 
study the effective Hamiltonian 
\bea  
H&=&H_{BCS}(1)+H_{BCS}(2) 
 - \varepsilon_J \sum_{j,k}^{\L}\left(b^{\d}_j(1)b_k(2)+
b^{\d}_j(2)b_k(1)\right), \label{jham} \eea    
 where $\varepsilon_J$ is the Josephson coupling energy, for the purpose 
of investigating the nature of pair tunneling at the nano-scale level.
In a previous work\cite{lzmg} 
it was shown that the above Hamiltonian is integrable 
for $\varepsilon_J=g$ for the case when $H_{BCS}(1),\,H_{BCS}(2)$ have identical
single electron energy levels. Below we will extend this construction 
to the case where $H_{BCS}(1),\,H_{BCS}(2)$ describe non-identical systems.

\section{A universal integrable system.}

First we introduce  the Lie algebra $su(2)$ with generators 
$S^+,\,S^-,\,S^z$ satisfying the commutation relations 
\beq [S^z,\,S^{\pm}]=\pm S^{\pm},~~~[S^+,\,S^-]=2S^z. 
\label{comm} \eeq 
The Casimir invariant, which commutes with each element of the algebra, 
has the form 
$$C=S^+S^-+S^-S^++2\left(S^z\right)^2. $$ 
Associated with the $su(2)$  
algebra there is a solution of the Yang-Baxter equation
in ${\rm End} V\otimes {\rm End V}\otimes su(2)$, where $V$ denotes 
a two-dimensional vector space. This solution reads\cite{fad}  
$$R_{12}(u-v)L_{1}(u)L_{2}(v)=L_{2}(v)L_{1}(u)R_{12}(u-v)$$ 
with 
\bea 
R(u)&=&I\otimes I +\frac{\eta}{u}\sum_{m,n}^2 e^m_n \otimes e^n_m 
, \nn \\
L(u)&=&I\otimes I + \frac{\eta}{u}\left(e^1_1\otimes S^z
-e^2_2\otimes S^z 
+e^1_2\otimes S^-+e_1^2\otimes S^+ \right) \nn 
\eea 
where $\{e^m_n\}$ are $2\times 2$ matrices with 1 in the
$(m,n)$ entry and zeroes eleswhere. 
Above, $I$ is the identity operator and 
$\eta$ is a scaling parameter for the rapidity variable $u$ which
plays an important role in the subsequent analysis. 
With this solution we construct the transfer matrix 
\beq t(u)={\rm tr}_0\left(G_0L_{0{\L}}(u-\e_{\L})...L_{01}(u-\e_1)\right) 
\label{tm} \eeq  
which is an element of the $\L$-fold tensor algebra of $su(2)$.
Above, 
${\rm tr}_0$ denotes the trace
taken over the auxiliary space and $G=\exp(\alpha\eta {\sigma})$ with
$\sigma={\rm diag}(1,\,-1)$. 
A consequence of the Yang-Baxter equation is that 
$[t(u), \,t(v)]=0$ 
for all values of the parameters $u$ and $v$, and independent of the 
representations of $su(2)$ in the tensor algebra. 
Defining 
$$T_j=\lim_{u\rightarrow \e_j}\frac{u-\e_j}{\eta^2} t(u) $$ 
for $j=1,2,...,{\L}$,
we may write in the {\it quasi-classical limit}  
$T_j=\tau_j+o(\eta) $ 
and it follows that 
$ [\tau_j,\,\tau_k]=0, ~ \forall\, j,\,k. $ 
Explicitly, these operators read 
\beq 
\tau_j=2\alpha S^z_j+\sum_{k\neq j}^{\L}\frac{\theta_{jk}}{\e_j-\e_k}
\label{cons} \eeq 
with 
$\theta=S^+\otimes S^-+S^-\otimes S^++2S^z\otimes S^z.$ 

We define a Hamiltonian through
\bea  
H&=&-\frac{1}{\a}\sum_j^{\L}\e_j\tau_j
+\frac{1}{4\a^3}\sum_{j,k}^{\L}\tau_j\tau_k 
+\frac{1}{2\a^2}\sum_j^{\L} \tau_j- \frac{1}{2\a}\sum_j^{\L}C_j 
\\ 
&=& -\sum_j^{\L}2\e_jS_j^z-\frac{1}{\a}\sum_{j,k}^{\L}S_j^-S_k^+.
\label{ham1} \eea       
The Hamiltonian is {\it universally} integrable since it is clear that 
$[H,\,\tau_j]=0,~~\forall j $ irrespective of the realizations of the 
$su(2)$ algebra in the tensor algebra.  

Realizing the $su(2)$ generators through the hard-core bosons; viz
\beq S_j^+=b_j, ~~~
S_j^-=b_j^{\dagger}, ~~~ 
S^z_j=\frac{1}{2}\left(I-n_{j}\right) \label{psr} \eeq  
one obtains (\ref{bcs}) (up to a constant) with $g=1/\a$ 
as shown by Zhou et al.\cite{zlmg}
and von Delft and Poghossian\cite{vp}. 

We now turn to applying (\ref{ham1}) for the study of two coupled BCS systems. 
To accomodate this, it is convenient to first consider three index sets 
$P_0,\,P_1,\,P_2$ such that individually the BCS Hamiltonians are expressible 
\bea 
H_{BCS}(i)=\sum_{j\in (P_0\cup P_i)}^{\L}\e_jn_j
-g\sum_{j,k\in (P_0\cup P_i)}^{\L}b_k^{\d}b_j. 
\nn \eea 
If the single particle energy $\e_j$ is common to both systems, then $j\in P_0$.
Hence it is meant to be understood that 
$\e_j\neq\e_k\neq \e_l\,\forall\, j\in P_1,\,
k\in P_2,\,l\in P_3$. 
In the case that $j\in P_0$, 
the local $su(2)$ operators are described by the 
tensor product of two 
pseudo-spin realisations acting on the four-dimensional tensor 
product space. 
We can now realise (\ref{ham1}) in terms of the hard-core boson representation 
(\ref{psr}) 
$$ S^+_j=b_j(i),~~S^-_j=b_j^{\dagger}(i),~~S^z_j=\frac12(I-n_j(i)) $$ 
for $j\in P_i,\, i=1,2$ whereas for $j\in P_0$ we take the tensor 
product representation
\bea S_j^+&=&b_j(1)+b_j(2) \nn \\
S_j^-&=&b^{\dagger}_j(1)+b^{\dagger}_j(2) \nn \\
S_j^z&=&I-\frac12\left(n_j(1)+n_j(2)\right).  \nn \eea 
Under this representation of (\ref{ham1}) we obtain (\ref{jham}) with 
$\varepsilon_J=g=1/\a$, 
establishing integrability at this value of the Josephson 
coupling energy. For the case when the index sets $P_1,\,P_2$ are both empty,
i.e., the two BCS systems are identical, this result was 
previously shown by Links 
et al.\cite{lzmg}. 

\section{The exact solution.}

In addition to proving integrability for 
$\varepsilon_J=g$, we can also obtain the exact solution from the Bethe 
ansatz. Below we will derive the energy eigenvalues for the Hamiltonian
(\ref{ham1}) in a very general context, which includes those of 
(\ref{jham}) with 
$\varepsilon_J=g$ as a particular case. 

For each index $k$ in the tensor algebra in which the transfer matrix acts, 
and accordingly in (\ref{ham1}),
suppose that we represent the $su(2)$ 
algebra through the irreducible representation with spin $s_k$. 
Thus $\{S^+_k,\,S^-_k,\,S^z_k\}$ act on a $(2s_k+1)$-dimensional space. 
Employing the standard method of the algebraic Bethe ansatz\cite{fad} 
gives that the eigenvalues of the transfer matrix
(\ref{tm}) take the form
\bea
\Lambda(u)&=&\exp(\alpha\eta)\prod_k^{\L}\frac{u-\e_k+\eta s_k}{u-\e_k}
\prod_j^M\frac{u-w_j-\eta}{u-w_j} \nn \\
&&~
+\exp(-\alpha\eta)\prod_k^{\L}\frac{u-\e_k-\eta s_k}{u-\e_k}
\prod_j^M\frac{u-w_j+\eta}{u-w_j}. \nn  
 \eea
Above, the parameters $w_j$ are required to satisfy the Bethe ansatz
equations
$$\exp(2\alpha\eta)\prod_k^{\L}\frac{w_l-\e_k+\eta s_k}
{w_l-\e_k-\eta s_k}
=-\prod_j^M\frac{w_l-w_j+\eta}{w_l-w_j-\eta} . $$

The eigenvalues of the conserved operators (\ref{cons}) are obtained
through the appropriate 
terms in the expansion of the transfer matrix eigenvalues  
in the parameter $\eta$. This yields the following result 
for the eigenvalues $\lambda_j$ of $\tau_j$ 
\beq \lambda_j= \left(2\a +\sum_{k\neq j}^{\L}\frac{2s_k}{\e_j-\e_k} 
-\sum_i^M 
\frac{2}{\e_j-v_i}\right)s_j \label{eig} \eeq  
such  that the parameters $v_j$ satisfy the coupled algebraic equations 
\beq 
2\a+ \sum_k^{\L}\frac{2s_k}{v_j-\e_k}
=\sum_{i\neq j}^M \frac{2}{v_j-v_i}.
\label{bae} \eeq  
Through (\ref{eig}) we can now determine the energy eigenvalues of 
(\ref{ham1}). It is useful to note the following identities 
\bea 
2\a \sum_j^{M}v_j+2\sum_j^M\sum_k^{\L}\frac{v_js_k}{v_j-\e_k}
&=&M(M-1) \nn \\
\a M+\sum_j^M\sum_k^{\L}\frac{s_k}{v_j-\e_k}&=&0 \nn \\
\sum_j^M\sum_k^{\L}\frac{v_js_k}{v_j-\e_k} -
\sum_j^M\sum_k^{\L}\frac{s_k\e_k}{v_j-\e_k}&=&M\sum_k^{\L}s_k. 
\nn\eea 
Employing the above it is deduced that 
\bea 
\sum_j^{\L}\lambda_j &=&2\a\sum_j^{\L}s_j-2\a M \nn \\
\sum_j^{\L}\e_j\lambda_j&=&2\a \sum_j^{\L}\e_j s_j+\sum_j^{\L}
\sum_{k\neq j}^{\L} 
s_js_k-2M\sum_k^{\L}s_k-2\a\sum_j^M v_j +M(M-1) \nn \eea 
which, combined with the eigenvalues $2s_j(s_j+1)$ for the Casimir invariants
$C_j$, yields the energy eigenvalues 
\beq E=2\sum_j^{M} v_j -2\sum_k^{\L}s_k \e_k. \label{nrg} \eeq  
{}From the above expression we see that the quasi-particle excitation
energies are given by twice the Bethe ansatz roots $\{v_j\}$ 
of (\ref{bae}). 

In order to specialise this result to (\ref{jham}) at integrable coupling,
it is useful to first make the following observation. For
$j\in P_0$, in which case the $su(2)$ algebra is realised via the tensor product
of two hard-core boson representations,
it is well known
that the representation space is completely reducible into triplet states
and a singlet state. Note however, that for the singlet state the $su(2)$
generators act trivially, and hence this state is blocked 
from scattering in analogy with
the blocking of single particle states discussed in the introduction. Hence the
$su(2)$ algebra will only act non-trivially on the triplet states.
In specialising (\ref{bae},\ref{nrg}) to the case of (\ref{jham}), we need 
only to 
set $s_j=1/2$ for $j\in P_1\cup P_2$ and $s_j=1$ for $j\in P_0$. 

\section{Conclusion} 

We have displayed the existence of a general 
class of integrable systems which 
includes the reduced BCS Hamiltonian and a model for strong Josephson tunneling 
between two reduced BCS systems. By deriving the models through the QISM 
we have also determined the exact solution via the Bethe ansatz. 
A further application of this 
approach is the computation of form factors and correlation 
functions\cite{zlmg,lzmg}. 
  
\section*{Acknowledgements}

Jon Links thanks Departamento de 
F\'{\i}sica 
Te\'orica de 
Universidad de Zaragoza
for hospitality and Prof. M.-L. Ge for the kind invitation to attend
the APCTP-Nankai Joint Symposium celebrating the 70th birthday of Prof.
F.Y. Wu. We thank the Australian Research Council and the Ministerio
de Educaci\'on y Cultura, Espa\~na for
financial support.



\begin{thebibliography}{99}
\bibitem{brt} C.T. Black, D.C. Ralph and M. Tinkham, {\it Phys. Rev. Lett.}
{\bf 76}, 688 (1996). 
\bibitem{rbt} D.C. Ralph, C.T. Black and M. Tinkham, {\it Phys. Rev. Lett.}
{\bf 78}, 4087 (1997). 
\bibitem{dr01} J. von Delft and D. C. Ralph, {\it Phys. Rep.} {\bf 345}, 61
(2001). 
\bibitem{vd} J. von Delft, {\it Ann. Phys.}  {\bf 10}, 219 (2001).
\bibitem{bcs} L.N. Cooper, J. Bardeen and J.R. Schrieffer, {\it Phys. Rev.} 
{\bf 108}, 1175 (1957). 
\bibitem{r63} R.W. Richardson, {\it Phys. Lett.}
  {\bf 3}, 277 (1963);
    {\bf 5}, 82 (1963); \\
     R.W. Richardson and N. Sherman, {\it Nucl. Phys.}
       {\bf 52}, 221 (1964);
	 {\bf 52}, 253 (1964).
\bibitem{crs} M.C. Cambiaggio, A.M.F. Rivas and M. Saraceno,
    {\it Nucl.  Phys.} {\bf A624}, 157 (1997).
\bibitem{zlmg} H.-Q. Zhou, J.
Links, R.H. McKenzie and M.D. Gould,
{\it Phys. Rev.}  {\bf B65}, 060502(R) (2002).
\bibitem{vp} J. von Delft and R. Poghossian, cond-mat/0106405. 
\bibitem{lzmg} J. Links, H.-Q. Zhou, R.H. McKenzie and M.D. Gould, 
cond-mat/0110105. 
\bibitem{j} B.D. Josephson, {\it Phys. Lett.} {\bf 1}, 251 (1962). 
\bibitem{fad} L.D. Faddeev, {\it Int. J. Mod. Phys.}  {\bf A10}, 1845 (1995).
\end{thebibliography}
\end{document}